\documentclass[twocolumn,superscriptaddress,longbibliography,
aps,prx,floatfix,preprintnumbers]{revtex4-2}

\usepackage[normalem]{ulem}
\usepackage{graphicx}
\usepackage{bm}
\usepackage{color}
\usepackage{epstopdf}
\usepackage{amsmath}
\usepackage{amssymb}
\usepackage{epstopdf}
\usepackage{lipsum}
\usepackage{nicefrac, xfrac}
\usepackage{gensymb}
\usepackage{multirow}
\usepackage{scrextend}
\usepackage[urlcolor=blue,colorlinks=true,citecolor=blue,linkcolor=blue,pdfstartview={FitH},bookmarks=false]{hyperref}
\usepackage[dvipsnames]{xcolor}
\usepackage{braket}

\graphicspath{{fig/}{./fig/}{.}}

\begin{document}

\title{
Theoretical investigation of the photovoltaic properties of MgSnN$_{2}$ \\ for multi-junction solar cells
}

\author{Issam~Mahraj}
\email[e-mail: ]{issam.mahraj@ifj.edu.pl}
\affiliation{Institute of Nuclear Physics, Polish Academy of Sciences, W. E. Radzikowskiego 152, PL-31342 Krak\'{o}w, Poland}

\author{Mossab~Oublal}
\email[e-mail: ]{mossab.oublal@edu.uiz.ac.ma}
\affiliation{Laboratory of Condensed Matter Physics and Nanomaterials for Renewable Energy, Faculty of Sciences, Ibn Zohr University, 80000 Agadir, Morocco}

\author{Andrzej~Ptok}
\email[e-mail: ]{aptok@mmj.pl}
\affiliation{Institute of Nuclear Physics, Polish Academy of Sciences, W. E. Radzikowskiego 152, PL-31342 Krak\'{o}w, Poland}

\date{\today}

\begin{abstract}
The orthorhombic crystal structure of the MgSnN$_2$ compound with $Pna2_1$ symmetry has been investigated as a low-cost, non-toxic material for photovoltaic (PV) applications using density functional theory (DFT) and spectroscopic limited maximum efficiency (SLME) calculations.
A detailed analysis of the electronic and optical properties was performed using the mBJ semilocal exchange functional. The bandgap of MgSnN$_2$ is found to be $2.45$~eV. 
SLME photovoltaic analysis suggests that a thin film of MgSnN$_2$ with a thickness of 2~$\mu$m can reach an efficiency of $13.17\%$ at room temperature.
This efficiency was further improved through the simulation  of a multi-junction device, where the tandem configuration increased the efficiency from 12.80$\%$ (single-junction) to 22.42$\%$.
Furthermore, introducing cation disorder can further reduce the bandgap, enhancing its suitability for solar cell applications.
\end{abstract}

\maketitle

\section{Introduction}

Heterovalent ternary nitride materials, characterized by the general formula II-IV-N$_{2}$ (where II includes Mg, Zn and Cd, and IV includes Si, Ge, and Sn), have recently garnered growing interest~\cite{punya2011, quirk14, lambrecht2016, rasander2018,le18, javaid18, makin2019, lambrecht2019, nicole20}.
They have been proposed as interesting candidate materials in, e.g., UV-optoelectronic devices and solar cells~\cite{greenaway2020, dumre2021, fahad22}.
These materials are seen as complementary to the well-studied III-N semiconductors~\cite{zakutayev2016}, offering enhanced flexibility for tuning various properties.
While the latter is well known for its wurtzite structure~\cite{zakutayev2016}, such chemical substitution, by replacing the group-III element by alternating group-II
and group-IV elements, is expected to transform the wurtzite structure into a so-called wurtzite-derived ordered superstructure, which
has the space group $Pna2_{1}$ structure~\cite{lambrecht2016}. 
Such this transformation is well known for other II-IV-V$_{2}$ semiconductors (here V other than N), where it changes the zincblende structure into the chalcopyrite structure~\cite{issam24}.
For example, this strategy has been successfully employed to develop ZnGeN$_{2}$ and ZnSnN$_{2}$~\cite{le18, javaid18, nicole20}.

\begin{figure}[b]
\centering
\includegraphics[width=\linewidth]{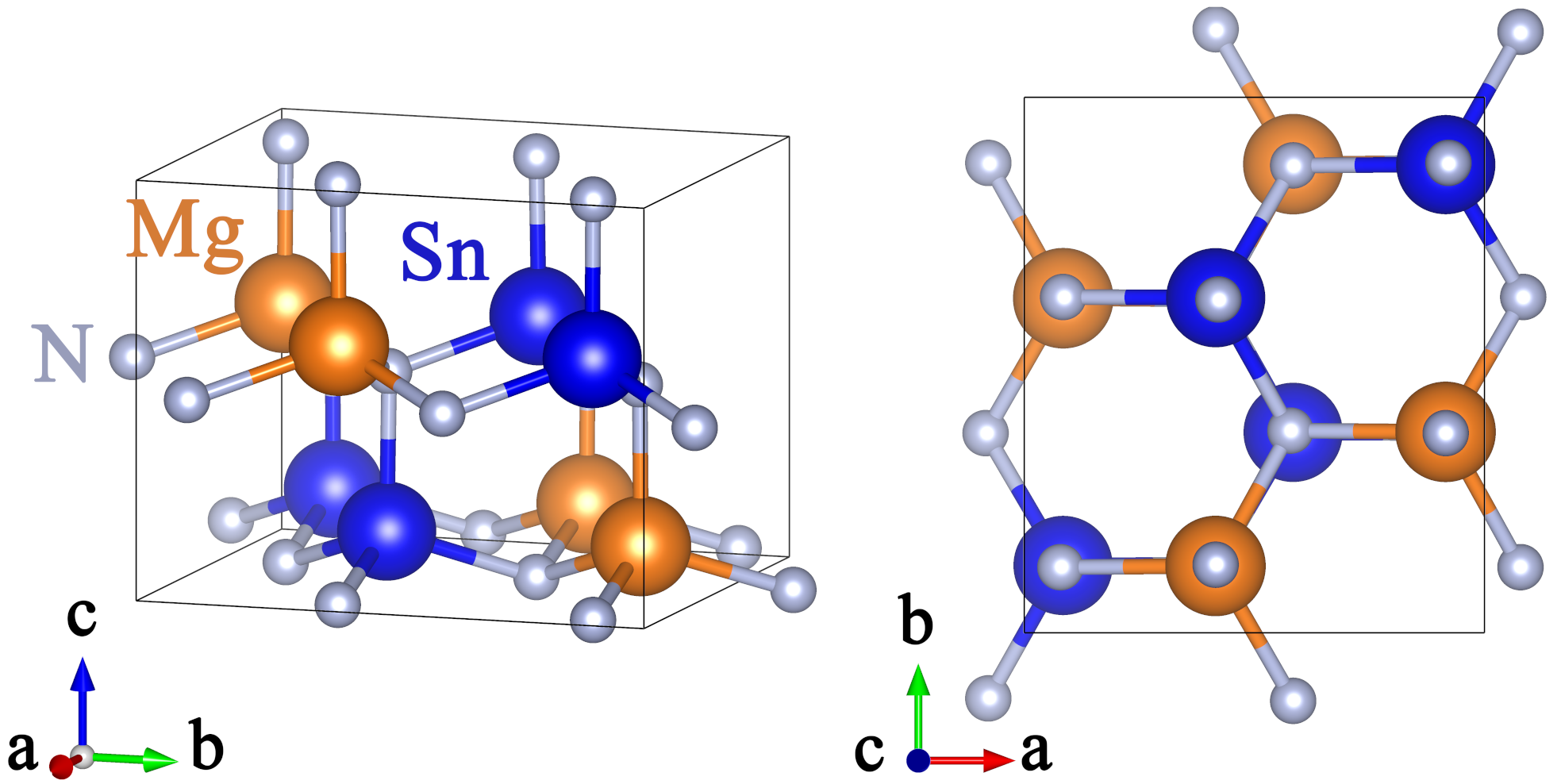}
\caption{
Standard view (left panel) and top view (right panel) of cation-ordered wurtzite-type of MgSnN$_{2}$ compound with $Pna2_{1}$ space group.
\label{str}
}
\end{figure}

The crystal structures and bonding properties of II-IV-N$_{2}$ semiconductors are related to those of III-N~\cite{lambrecht2016,rasander2018}.
However, they offer different combinations of bandgaps and lattice parameters~\cite{punya2011, lambrecht2016}, thereby opening up additional possibilities for device design, suggesting either as complements to or as replacements for III-N materials.
For example: wide bandgap II-IV-N$_{2}$ compounds such as MgSiN$_{2}$ and MgGeN$_{2}$ may find applications in UV optoelectronic devices~\cite{lambrecht2016, nicole20}, while materials like ZnSnN$_{2}$ could be used in solar cells~\cite{le18, javaid18}.
However, before new materials can be useful for applications, it is necessary to establish their fundamental physical properties.
For this reason, we found it necessary to study the physical properties, especially of the MgSnN$_{2}$ compound, due to the lack of studies available in the literature.

The crystal structures of MgSiN$_{2}$ and MgSnN$_{2}$ have been confirmed through X-ray diffraction analysis to crystallize in the orthorhombic $Pna2_{1}$ space group (Fig.~\ref{str})~\cite{quirk14, lambrecht2016}.
However, the precise crystal structure of MgSnN$_{2}$ remains a subject of ongoing investigation.
Alternative structural models, such as a disordered wurtzite-like configuration with the $Pmc2_{1}$ space group, have also been proposed~\cite{quayle15}.
Cation-disordered wurtzite-type structures with secondary rocksalt-type phases in Mg-rich samples at low
temperatures have been observed~\cite{greenaway2020}.
Furthermore, high-pressure metathesis reactions have yielded a disordered rocksalt-like structure~\cite{kawamura20}.
Additionally, epitaxial growth techniques have facilitated the formation of MgSnN$_{2}$ films with a hexagonal ZnO-like structure~\cite{yamada21}.
More recently, a hexagonal structure with the $P6_{3}mc$ space group has been reported for MgSnN$_{2}$~\cite{fahad22}.
However, good agreement is found between the measured absorption and the calculated spectra using the space group $Pna2_{1}$~\cite{fahad22}.
Notably, the bandgap energy has been found to vary from $1.87$ to $3.43$~eV depending on the long-range order parameter of the cation sublattice~\cite{makin2019}.

Conversely, theoretical investigations indicate that MgSnN$_{2}$ is most stable and energetically favorable in the $Pna2_{1}$ space group compared to alternative structures~\cite{greenaway2020, dumre2021}. 
The structural properties and electronic band structure of the wurtzite-based $Pna2_{1}$ phase have been explored via the quasiparticle self-consistent $GW$ approximation~\cite{lambrecht2016, lambrecht2019}. 
Furthermore, the elastic constants for this crystal structure have been determined through density functional theory (DFT) calculations~\cite{rasander2018}.

The current work focuses on heterovalent ternary nitride material MgSnN$_{2}$ for multi-junction solar cell and optoelectronic, which are environmentally friendly, low-cost, earth-abundant, and non-toxic.
These materials are composed of widely available elements, are significantly less expensive than In and Ga, and benefit from a well-established recycling infrastructure~\cite{makin2019}.
However, the efficient solar devices depend not only on electronic and optical parameters but also on the type of bandgap and critical solar cell parameters.
Some successful theoretical studies, using first-principles approaches, have identified new materials with high power conversion efficiency (PCE) for PV applications~\cite{yu2012, lee2014, yin2014}. 
One notable method for predicting the maximum efficiency of a single-junction solar cell is the Spectroscopic Limited Maximum Efficiency (SLME)~\cite{yu2012} which builds upon the Shockley--Queisser (SQ) limiting efficiency~\cite{shockley1961}.
SLME requires information about the bandgap value, absorption spectra, and recombination characteristics. 

This paper presents a detailed first-principles study using modified Becke-Johnson (mBJ) semilocal exchange functional~\cite{becke.johnson.06,tran.blaha.09} to evaluate the electronic, optical, and solar cell properties of MgSnN$_{2}$.
The mBJ method demonstrates significant potential for accurately predicting the bandgaps of semiconductors and insulators, offering a more efficient alternative to computationally intensive approaches such as $GW$ and HSE06~\cite{blaha.schwarz.20,issam24}.
The bandgap predicted by mBJ is $2.45$~eV.
The density of states reveals the presence of $p$-$p$ coupling between N $p$-states and Sn $p$-states at the valence band maximum (VBM), while the conduction band minimum (CBM) consists of a mix of Sn $s$-states and N $p$-states, leading to $sp^3$ hybridization in tetrahedral coordination.

Additionally, the static dielectric constant for MgSnN$_{2}$ is estimated to be approximately $4.05$ resulting in a refractive index of~$2.01$. 
MgSnN$_{2}$ exhibits a high absorption coefficient of around $10^{5}$~cm$^{-1}$ within the active solar spectrum. 
However, high absorption alone does not guarantee an efficient single-junction solar cell.
Therefore, SLME for an ideal pn-junction solar cell is calculated as a function of thickness.
The maximum possible efficiency is estimated to be $13.17\%$ at room temperature for a thickness of 2~$\mu$m.

To evaluate the potential of MgSnN$_2$ for multi-junction applications, we first fabricated a single-junction device to validate the SLME results.
The fabricated device achieved an efficiency of $12.80\%$ at room temperature, closely aligning with SLME value.
Subsequently, a multi-junction device was simulated by connecting two sub-cells in series.
The simulation demonstrated notable enhancements in key performance parameters, including open-circuit voltage, maximum voltage, and power density.
The overall efficiency increased significantly—from 12.80$\%$ in the single-junction configuration to 22.42$\%$ in the tandem structure.

Finally, MgSnN$_{2}$, with its wide bandgap and high absorption coefficient, is proposed as a promising material for the top layer in multi-junction solar devices. 
Introducing cation disorder can further reduce the bandgap~\cite{makin2019} and enhance low-energy absorption~\cite{greenaway2020}, thereby improving the material's suitability for solar cell applications.

The structure of the paper is as follows: Section~\ref{sec.comp} provides details of the numerical calculations.
Our results are presented and analyzed in Section~\ref{sec.results}. Finally, Section~\ref{sec.sum} summarizes our findings and offers concluding remarks.

\section{Methods}
\label{sec.comp}

{\it Electronic properties }---
To study the electronic properties, density functional theory (DFT) calculations were performed using the {\sc Wien2}k code~\cite{blaha.schwarz.20}, based on the FP-LAPW method. A plane wave cutoff of $R_{MT} \times K_{max} = 7$ was used, with muffin-tin radii of 1.93, 2.07, and 1.70 a.u. for Mg, Sn, and N, respectively. Total energy was minimized using both LDA~\cite{perdew.wang.92} and GGA-PBE~\cite{perdew.burke.96} functionals.
Since LDA and GGA tend to underestimate the bandgap~\cite{issam24}, the modified Becke–Johnson (mBJ) potential~\cite{becke.johnson.06,tran.blaha.09} was employed for improved accuracy~\cite{blaha.schwarz.20,issam24}.
The first Brillouin zone was sampled using a $17 \times 17 \times 17$ ${\bm k}$-grid and a $21 \times 21 \times 21$ ${\bm k}$-grid in the Monkhorst–Pack scheme~\cite{monkhorst.pack.76} for volume optimization and self-consistent field (SCF) calculations, respectively.
The total energy in the SCF iterations converged to within $10^{-4}$~Ry.

{\it Optical properties }---
For the optical properties, we calculated the complex dielectric function, $\epsilon(\omega) = \epsilon_{1}(\omega) + i\epsilon_{2}(\omega)$, in the continuum region above the quasiparticle gap using the independent particle approximation (IPA).
This function describes the optical response of a material to incident light with photon energies $\hbar\omega$.
The imaginary part, $\epsilon_{2}(\omega)$, in the long-wavelength limit, was directly derived from electronic structure calculations using the following expression~\cite{delin98}:
\begin{eqnarray}
\nonumber 
\epsilon_{2}(\omega) = \frac{4\pi^{2}e^{2}}{m^{2}\omega^{2}} \sum_{i,j} \int \braket{j|M|i}^{2}f_{j}(1-f_{i})\delta(\Delta E-\hbar\omega)d^{3}k . 
\\ \label{ep2}
\end{eqnarray}
Here, $M$, $i$, and $j$ denote the dipole matrix element, the initial state, and the final state, respectively.
$f_{i}$ is the Fermi distribution function for the $i$th state, and $\Delta E = E_{j} - E_{i}$ represents the energy difference between the $i$th and $j$th states.
The real part of the dielectric function is obtained from the imaginary part using the Kramers–Kronig relation, expressed as~\cite{Kronig.26}:
\begin{equation}
\label{ep1}
\epsilon_{1}(\omega) = 1 + \frac{2}{\pi} P \int \frac{\omega'\epsilon_{2}(\omega')}{\omega'^{2}-\omega^{2}} d\omega' .
\end{equation}
Here, $P$ denotes the principal value of the integral.
The absorption coefficient, $\alpha(\omega)$, the refractive index, $n(\omega)$, and the optical reflectivity, $r(\omega)$, are calculated using the following expressions:
\begin{eqnarray}
\label{alph}
\alpha(\omega) &=& \frac{2\omega k(\omega)}{c} , \\
\label{n}
n(\omega) &=& \frac{1}{\sqrt{2}}\left[2\sqrt{\epsilon^{2}_{1}(\omega)+\epsilon^{2}_{2}(\omega)}+\epsilon_{1}(\omega)\right]^{1/2} , \\
\label{R}
r(\omega) &=& \frac{(n(\omega)-1)^{2}+k^{2}(\omega)}{(n(\omega)+1)^{2}+k^{2}(\omega)} ,
\end{eqnarray}
where $k(\omega)$ is the coefficient of extinction, obtained from:
\begin{equation}
\label{k}
k(\omega) = \frac{1}{\sqrt{2}}\left[\sqrt{\epsilon^{2}_{1}(\omega)+\epsilon^{2}_{2}(\omega)}-\epsilon_{1}(\omega)\right]^{1/2} .
\end{equation}

{\it Photovoltaic properties }---
The photovoltaic performance is evaluated through absorption-based first-principles calculations using the spectroscopic limited maximum efficiency (SLME) model~\cite{yu2012}.
This model employs the standard solar spectrum, $I_i$, under standard test conditions (STC), specifically AM1.5G and $T = 25^{\circ}$C~\cite{stc}.
By utilizing the calculated absorption coefficient from the mBJ potential and the material's thickness, the SLME model predicts the maximum achievable solar efficiency.
The theoretical efficiency is defined as follows~\cite{yu2012}:
\begin{equation}
\label{TE}
\eta = \frac{P_{max}}{P_{in}} ,
\end{equation}
where $P_{max}$ is the maximum output power density, obtained by numerically optimizing the product of current density and voltage, and $P_{in}$ is the total incident solar power density, equal to $1000$~W/m$^{2}$.
The total current density of an ideal solar cell junction under photon flux is expressed as:
\begin{equation}
\label{J}
J = J_{sc} - J_{0} \left\{ \exp \left( \frac{qV}{k_{B}T} \right) - 1 \right\} .
\end{equation}
Thus, the maximum power density for the SLME is estimated using the solar cell's $J$--$V$ characteristics:
\begin{equation}
\label{Pmax}
P_{max} = \text{max} \left\{ \left[ J_{sc} - J_{0} \left( \exp \left(\frac{qV}{k_{B}T} \right) - 1 \right) \right] V \right\} .
\end{equation}
Here, $q$, $T$, $k_{B}$, and $V$ denote the elementary charge of an electron, the temperature, the Boltzmann constant, and the voltage, respectively.
The short-circuit current density, $J_{sc}$, and the recombination current density (or reverse saturation current), $J_{0}$, are defined as follows:
\begin{eqnarray}
\label{Jsc}
J_{sc} &=& q\hbar\int a(\omega)I_{i}(\omega)d\omega , \\
\label{J0}
J_{0} &=& J_{0}^{r} + J_{0}^{nr} = \frac{J_{0}^{r}}{f_{r}},
\end{eqnarray}
Here, $J_{0}^{r}$ and $J_{0}^{nr}$ represent the total radiative and nonradiative electron-hole recombination currents at equilibrium under dark conditions, respectively, while $f_{r}$ denotes the radiative recombination fraction, defined as:
\begin{equation}
\label{fr}
f_{r} =\exp \left( \frac{E_{g}^{opt} - E_{g}}{k_{B}T} \right) .
\end{equation}
Here, $E_{g}^{opt}$ is the optical bandgap (direct bandgap) and $E_{g}$ is the fundamental bandgap (minimum bandgap) of the material. 

Under equilibrium conditions, the rate of photon emission due to radiative recombination equals the rate of photon absorption through the cell surface from the surrounding medium in the dark.
Assuming the cell is surrounded by an ideal heat sink and that the surrounding temperature matches the solar cell's temperature, the absorption of black-body photons at temperature, $T$, by the cell's front surface from the thermal bath determines $J_{0}^{r}$ as:
\begin{equation}
\label{J0r}
J_{0}^{r} = q\pi\hbar \int a(\omega)I_{0}(\omega)d\omega ,
\end{equation}
where, $I_{0}$ represents the blackbody spectrum, and $a(\omega)$ is the absorptivity.
Thus, the recombination current density is given by:
\begin{equation}
\label{J02}
J_{0} = \frac{q\pi\hbar}{f_{r}} \int a(\omega)I_{0}(\omega)d\omega .
\end{equation}
In the Shockley-Queisser model~\cite{shockley1961}, only radiative recombination is considered, with $f_{r} = 1$.
Here, the photon absorptivity, $a(\omega)$, is given by:
\begin{equation}
\label{a}
a(\omega) = 1 - \exp \left(- f_{g}\alpha(\omega)x \right) .
\end{equation}
The conditions are defined such that $a(\omega) = 1$ for $E \geq E_{g}$ and $a(\omega) = 0$ for $E < E_{g}$.
$f_{g}$ represents the effect of the solar cell geometry on efficiency within the SLME model.
$\alpha(\omega)$ is the absorption coefficient obtained from DFT calculations, and $x$ is the thickness of the thin-film layer.
Additionally, we assume $f_{g} = 1$, indicating that the configuration includes a perfect reflector on the back side of the solar cell to prevent light from being emitted from the rear.

Finally, the open-circuit voltage, $V_{oc}$, which corresponds to the device's voltage under zero current density conditions, is determined as follows:
\begin{equation}
\label{Voc}
V_{oc} = \frac{k_{B}T}{q} \ln \left( 1 + \frac{J_{sc}}{J_{0}} \right) ,
\end{equation}
and the fill factor $FF$ is defined as:
\begin{equation}
\label{FF}
FF = \frac{P_{max}}{J_{sc}V_{oc}} .
\end{equation}

\section{Results and discussion}
\label{sec.results}

\subsection{Crystal structure}

The orthorhombic crystal structure of the MgSnN$_{2}$ compound, with $Pna2_{1}$ symmetry (space group No.33), is shown in Fig.\ref{str}.
This structure can be derived from the wurtzite structure by replacing every two group III atoms with one group II atom and one group IV atom.
The transformation is described by the relations $a = a_{1} + 2a_{2}$ and $b = 2a_{2}$, where $a_{1}$ and $a_{2}$ are the lattice constants of the wurtzite structure in the $xy$-plane, while $a$ and $b$ are the lattice vectors in the $xy$-plane of the orthorhombic structure.
The $c$ lattice vector is common to both crystal structures.
In Table~\ref{tab1}, we present the calculated lattice constants $a$, $b$, and $c$ for the orthorhombic MgSnN$_{2}$ system using LDA and GGA-PBE.
Our results are in good agreement with previous theoretical calculations from Refs.~\cite{lambrecht2016, rasander2018}.

Since the orthorhombic structures are derived from the wurtzite structure, we define a wurtzite-like lattice constant in the $xy$-plane for the orthorhombic structures as $\bar{a}_{w} = (a/\sqrt{3} + b/2)/2$, and the deviation from hexagonal symmetry as $\Delta_{w} = |a/\sqrt{3} - b/2|/\bar{a}_{w}$.
Based on our calculations, the average wurtzite-like lattice constant, $\bar{a}_{w}$, is found to be $3.392$~\AA\ and $3.460$~\AA\ using LDA and GGA-PBE, respectively, while the deviation from hexagonality, $\Delta_{w}$, is $0.71$\% using both methods.
These values are in good agreement with those obtained using PBEsol, where $\bar{a}_{w} = 3.425$~\AA\ and $\Delta_{w} = 0.8$\%~\cite{rasander2018}.

\begin{table}[b]
\caption{\label{tab1}
The calculated lattice parameters $a$, $b$, and $c$ (\AA), lattice volume $V$ (\AA$^{3}$), the average wurtzite-like lattice constant $\bar{a}$ (\AA), the deviation from hexagonality $\Delta_{w}$, and Wyckoff ($4a$) positions (reduced coordinates) in the unit cell of MgSnN$_{2}$ compound.
}
\begin{ruledtabular}
\begin{tabular}{cccc}
\textrm{}&
\textrm{LDA}&
\textrm{GGA-PBE}\\
\hline
$a$ & 5.8545 & 5.9728 \\
$b$ & 6.8086 & 6.9461 \\
$c$ & 5.4195 & 5.5290 \\
$V$ & 216.0264 & 229.3853 \\
$\bar{a}_{w}$ & 3.392 & 3.460 \\
$\Delta_{w}$ & 0.713 & 0.712 \\
Mg ($x,y,z$) & (0.084,0.624,0.000) & (0.083,0.624,0.000) \\
Sn ($x,y,z$) & (0.084,0.125,0.002) & (0.083,0.125,0.002) \\
N$_{Sn}$ ($x,y,z$) & (0.078,0.122,0.381) & (0.078,0.122,0.380) \\
N$_{Mg}$ ($x,y,z$) & (0.084,0.627,0.383) & (0.085,0.627,0.381) \\
\end{tabular}
\end{ruledtabular}
\end{table}

The crystal structures of related compounds, such as MgSiN$_{2}$ and MgGeN$_{2}$, are found to belong to the space group $Pna2_{1}$.
In this structure, the atoms Mg, Sn, N$_{Mg}$ (above Mg), and N$_{Sn}$ (above Sn) each occupy Wyckoff ($4a$) positions with the following reduced coordinates:
($x, y, z$), ($-x, -y, z + 1/2$), ($x + 1/2, -y + 1/2, z$), and ($-x + 1/2, y + 1/2, z + 1/2$), respectively.
The origin is chosen to lie on the two-fold screw axes parallel to the $c$-axis.
The $z$-coordinate of the origin is arbitrary, and we select it such that the $z$ positions of Mg are set to zero.
The parameters describing the reduced coordinates ($x, y, z$) for each atom type are listed in Table~\ref{tab1}.

We also determined the equilibrium bulk modulus, $B$, and its first pressure derivative, $B'$, by fitting the total energy as a function of volume to the Birch--Murnaghan and Rose--Vinet equations of state, using Eq.(\ref{BM}) and Eq.(\ref{VR}), respectively.
\begin{eqnarray}
\nonumber 
E &=& E_{0} + \frac{9BV_{0}}{16}\left[B'(\eta^{2}-1)^{3} + (\eta^{2}-1)^{2}(6-4\eta^{2}) \right] , \\
\label{BM} \\
\label{VR} E &=& E_{0}+\frac{4BV_{0}}{B'_{2}}
-\frac{2BV_{0}}{B'_{2}} \\
\nonumber && \times \left[5+3B'(\eta-1)-3\eta \right] \exp(-1.5B'(\eta-1)) .
\end{eqnarray}
Here, $\eta = (V_{0}/V)^{1/3}$, where $V_{0}$ and $E_{0}$ represent the optimized unit cell volume and the ground-state energy, respectively.
The obtained values are presented in Table~\ref{tab2}.
The reduced Wyckoff ($4a$) coordinates, bulk moduli, and their first pressure derivatives are consistent with those reported in Ref.~\cite{lambrecht2016}.

\begin{table*}
\caption{\label{tab2} 
The obtained $V_{0}$, $E_{0}$, Bulk moduli B, and their pressure derivatives B' of MgSnN$_{2}$ compound from fit to Birch–Murnaghan (BM) and Rose–Vinet (RV) equations of state.}
\begin{ruledtabular}
\begin{tabular}{cccccccccc}
 &\multicolumn{4}{c}{LDA}&\multicolumn{4}{c}{GGA} \\
 & $V_{0}$ (\AA$^{3}$) & $E_{0}$ ($R_{y}$) & $B $(GPa) & $B'$ & $V_{0}$ (\AA$^{3}$) & $E_{0}$ ($R_{y}$) & $B$ (GPa) & $B'$ \\ 
\hline
BM & 215.32 & 51857 & 92.9403 & 3.8586 & 235.98 & 51912 & 81.5331 & 3.9706 \\
RV & 215.30 & 51857 & 92.3413 & 3.6368 & 235.94 & 51912 & 81.9284 & 3.9697 \\
\end{tabular}
\end{ruledtabular}
\end{table*}

\subsection{Electronic properties}

\begin{figure}[!b]
\centering
\includegraphics[width=\linewidth]{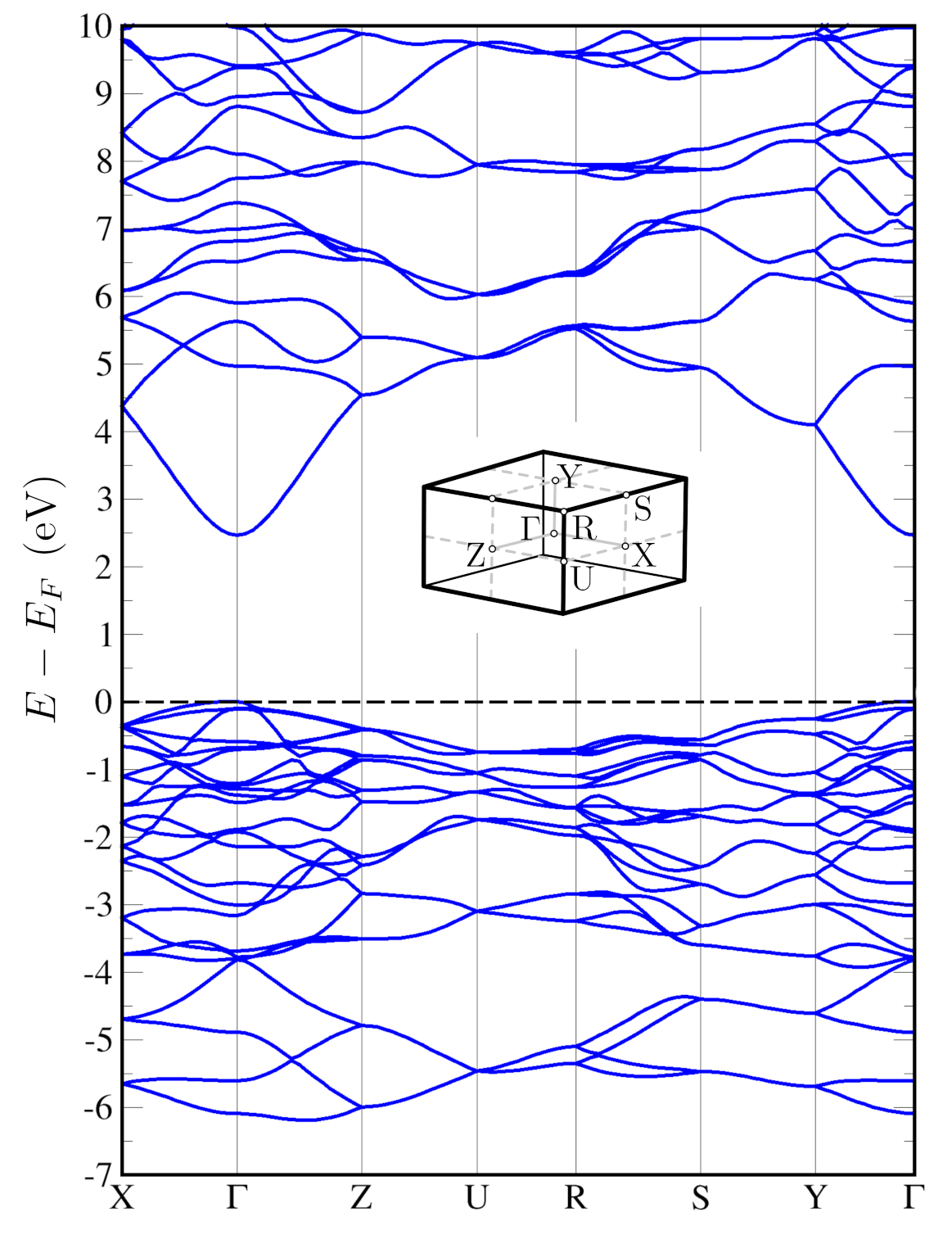}
\caption{The electronic band structure of MgSnN$_{2}$ along high symmetry direction of the Brillouin zone (presented in inset). Results obtained from the mBJ method.
\label{BS}
}
\end{figure}

The electronic band structure of MgSnN$_{2}$, calculated using the mBJ method at the LDA lattice constants, is shown in Fig.~\ref{BS}.  
The energy scale is referenced to the Fermi energy.  
As shown in Fig.~\ref{BS}, MgSnN$_{2}$ is a direct bandgap semiconductor, with both the valence band maximum (VBM) and conduction band minimum (CBM) occurring at the $\Gamma$ point.  
The conduction band edge consists of a single band, while the valence band edge features multiple bands due to crystal-field splitting.  
At the CBM, the electronic dispersion is nearly symmetric along the $x$, $y$, and $z$ directions, with low dispersion, suggesting small and isotropic electron effective masses, which are favorable for efficient electron transport.  
In contrast, at the VBM, the electronic dispersion is larger and asymmetric across these directions, indicating large and anisotropic hole effective masses.  
Specifically, the hole effective masses at the VBM are larger in the $x$ and $y$ directions compared to the $z$ direction.  
This observation is consistent with the effective masses calculated using $GW$ for both the VBM and CBM states~\cite{lambrecht2016, lambrecht2019}.  
The large hole effective masses at the VBM could also be beneficial for thermoelectric performance.

\begin{table}[b]
\caption{\label{tab3}
The calculated bandgap energy (in eV) of MgSnN$_{2}$ compound in various approximations with space group $Pna2_{1}$ symmetry.
}
\begin{ruledtabular}
\begin{tabular}{ccccc}
LDA & GGA & mBJ & {\it GW} & Exp. \\
\hline
1.26 & 1.31 & 2.45 & 2.47~\cite{greenaway2020} & 1.87-3.43~\cite{makin2019}\\
 & 1.6~\cite{lambrecht2016} & 2.34~\cite{rasander2018} & 3.43~\cite{lambrecht2016}\footnote{without semi-core states} & 2.27-2.43~\cite{rasander2018}\\
 & 1.17~\cite{lambrecht2019} & & 2.28~\cite{lambrecht2019}\footnote{with semi-core states} & \\
\end{tabular}
\end{ruledtabular}
\end{table}

\begin{figure}[!b]
\centering
\includegraphics[width=\linewidth]{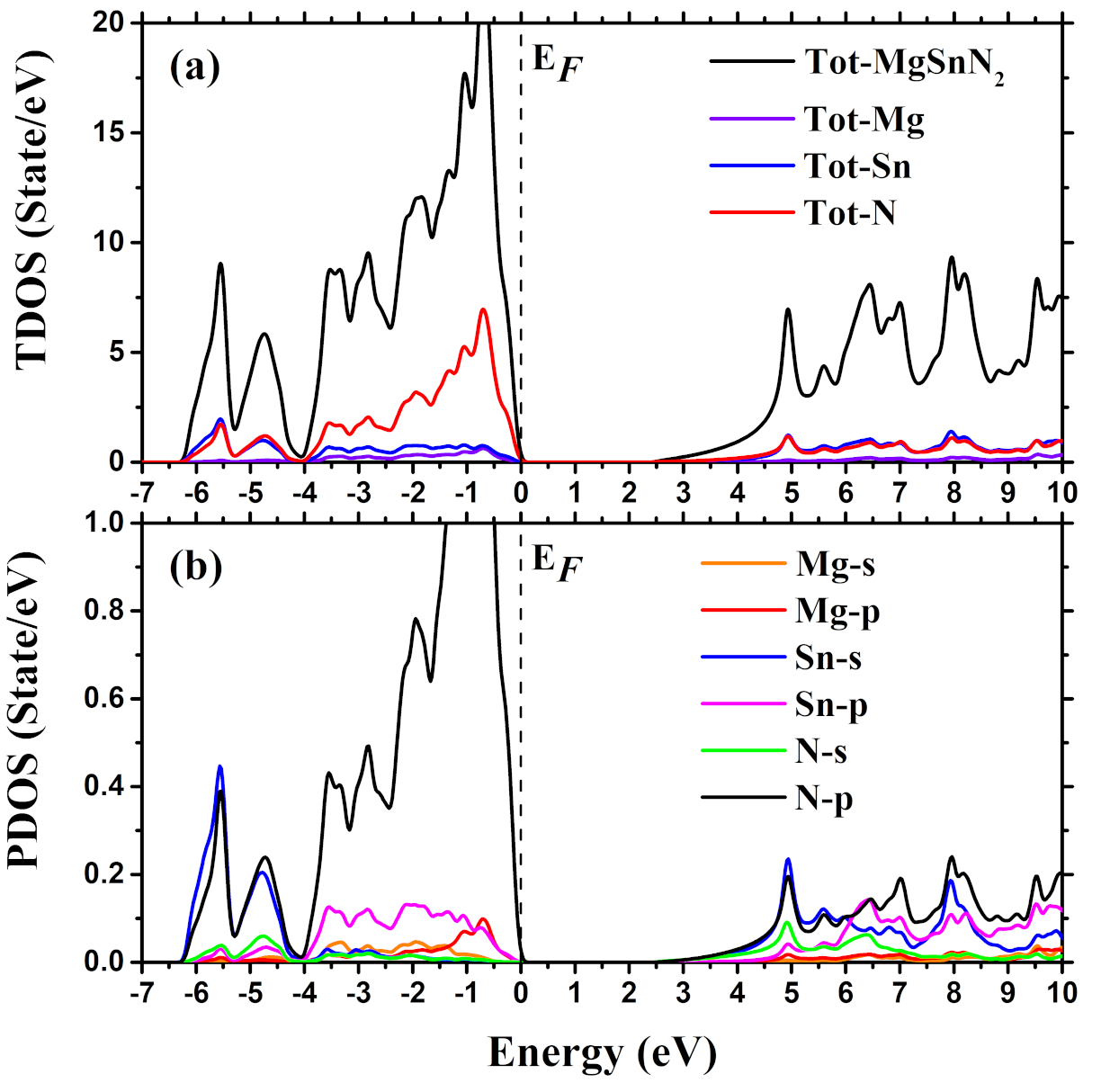}
\caption{
(a) Total densities of states (TDOS) and (b) partial densities of states (PDOS) of MgSnN$_{2}$.
\label{DOS}
}
\end{figure}

The bandgaps, calculated using various approximations, are provided in Table~\ref{tab3}.
As mentioned earlier, both LDA and GGA tend to underestimate the true bandgap compared to experimental results~\cite{issam24}.
For this reason, we employed the mBJ method.
It is important to note that all of our bandgap calculations were performed using the LDA lattice constants, which may result in larger values compared to those calculated using the GGA lattice constants~\cite{lambrecht2016, lambrecht2019}.
This discrepancy arises due to the negative bandgap deformation potential of MgSnN$_{2}$~\cite{lambrecht2016, lambrecht2019}.
For example, our GGA calculation yields a bandgap of $1.31$~eV, while Ref.~\cite{lambrecht2016} reports a value of $1.16$~eV using the GGA lattice constants.
The mBJ method produces a bandgap of $2.45$~eV, which is slightly smaller than the $3.59$~eV reported using the $GW$ method~\cite{lambrecht2016}.
This is expected, as the $GW$ method tends to systematically overestimate bandgaps due to its underestimation of dielectric screening~\cite{lambrecht2016}.
However, including $d$-states reduces the bandgap to $2.28$~eV~\cite{lambrecht2019}, which is in better agreement with our results.

The total and partial densities of states corresponding to the calculated mBJ electronic band structure are shown in Fig.~\ref{DOS}.
The partial density of states (PDOS) is summed over all equivalent atoms in the unit cell, including the four Mg atoms, four Sn atoms, and eight N atoms.
Note that the PDOS does not add up to the total density of states (TDOS) because the TDOS also includes contributions from the interstitial regions.
It is clear that the TDOS in the valence band is higher than in the conduction band, especially near the VBM, which results from the large electronic dispersion at VBM.
The Fermi level crosses the valence band, indicating that MgSnN$_{2}$ is a $p$-type semiconductor.
The VBM is primarily composed of N states, with a small contribution from Sn states.
According to the PDOS, these are mainly N $p$-states, with a smaller contribution from Sn $p$-states.
The Sn $s$-states make their largest contribution to the bonding states in the lower-energy region of the upper valence band, around $-6$~eV.
These results suggest a $p$-$p$ coupling between N $p$-states and Sn $p$-states in the VBM.
At the CBM, the TDOS is composed equally of Sn and N states.
The PDOS indicates that these states are $s$- and $p$-states of N atoms, along with $s$-states from Sn atoms.
Thus, the CBM consists of a mixture of $s$- and $p$-states, leading to $sp^3$ hybridization in tetrahedral coordination.

\subsection{Optical properties}

The performance and efficiency of any solar device are heavily influenced by the optical properties of the material.
These properties are primarily determined through response function calculations, which are expressed in terms of the frequency-dependent complex dielectric function, $\epsilon(\omega)$.

The dielectric function, $\epsilon(\omega)$, consists of two main contributions: intra-band and inter-band transitions.
In metals, the contribution from intra-band transitions is significant, while inter-band transitions can be further categorized as direct and indirect.
Indirect inter-band transitions, which involve phonon scattering and are expected to have a minimal effect on $\epsilon(\omega)$, are neglected.
To determine the contribution of direct inter-band transitions to the imaginary part of the dielectric function, $\epsilon_{2}(\omega)$, one sums over all possible electronic transitions from filled to empty states, as described by Eq.(\ref{ep2}).
The real part of the dielectric function, $\epsilon_{1}(\omega)$, is then obtained using the Kramers–Kronig relation in Eq.(\ref{ep1}).

\begin{figure}[!t]
\centering
\includegraphics[width=\linewidth]{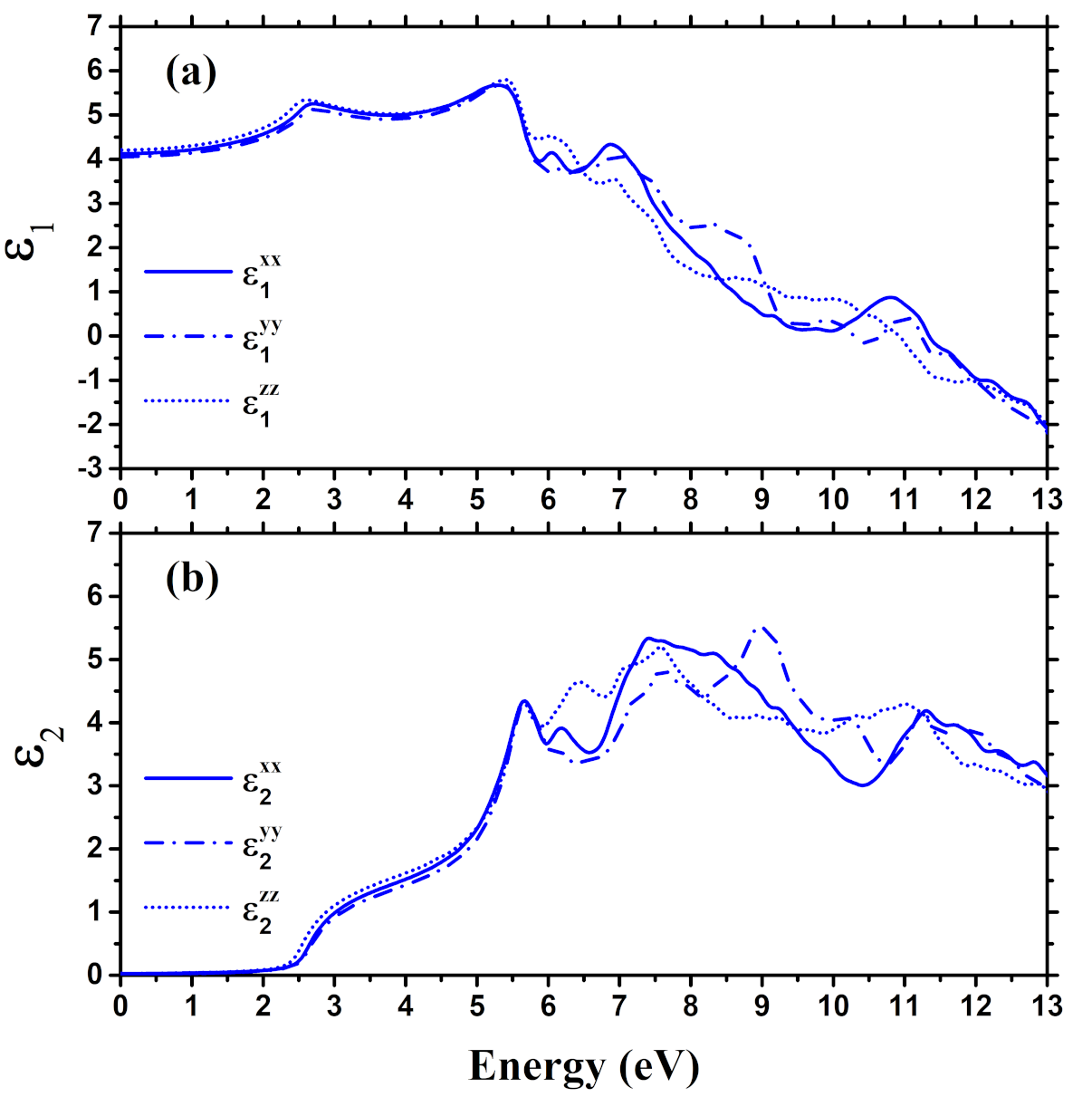}
\caption{
Dielectric function of the MgSnN$_{2}$ compound as a function of energy for the means diagonal components of the dielectric tensor: (a) real part, $\epsilon{1}(\omega)$, and (b) imaginary part, $\epsilon_{2}(\omega)$.
\label{epsilon}
}
\end{figure}

The calculated values of $\epsilon_{1}(\omega)$ and $\epsilon_{2}(\omega)$ for the three principal diagonal components are shown in Fig.~\ref{epsilon}.
Both $\epsilon_{1}(\omega)$ and $\epsilon_{2}(\omega)$ exhibit isotropic behavior in the low-energy range, but they display anisotropic behavior at higher energies.
For $\epsilon_{2}(\omega)$, which is closely related to the electronic band structure, we used the mBJ band structure in our calculations.
$\epsilon_{2}(\omega)$ is a crucial parameter that characterizes the optical response of a material and significantly influences its absorption properties.
It starts at the fundamental bandgap energy.
The first peak in the $\epsilon_{2}(\omega)$ spectrum appears in the visible region, suggesting that MgSnN$_{2}$ could be a promising material for photovoltaic applications.
However, the highest peak occurs in the lower part of the ultraviolet (UV) range, indicating that MgSnN$_{2}$ may also have potential for applications in the UV range.

For $\epsilon_{1}(\omega)$, the static value is approximately $4.05$, which corresponds to the refractive index of the material.
Thus, the static refractive index, $n(0)$, can be estimated as the square root of the zero-frequency limit of $\epsilon_{1}(\omega)$, yielding $n(0) = 2.01$.
As shown in Fig.~\ref{epsilon}(b), there are four main peaks in the $\epsilon_{1}(\omega)$ spectrum.
These peaks, resulting from direct transitions, can be explained by the DOS.
They correspond to direct inter-band transitions from the valence band to empty states in the conduction band.
The first and second peaks, occurring at $2.5$~eV and $5.5$~eV, respectively, can be attributed to transitions from the N $p$-states of the VBM to the Sn $s$-states/N $p$-states at the CBM.
The third and fourth peaks, located around $7$~eV and $11$~eV, respectively, are due to transitions from Sn $s$-states/N $p$-states in the valence band to Sn $s$-states/N $p$-states in the conduction band.

\begin{figure*}
\centering
\includegraphics[width=\linewidth]{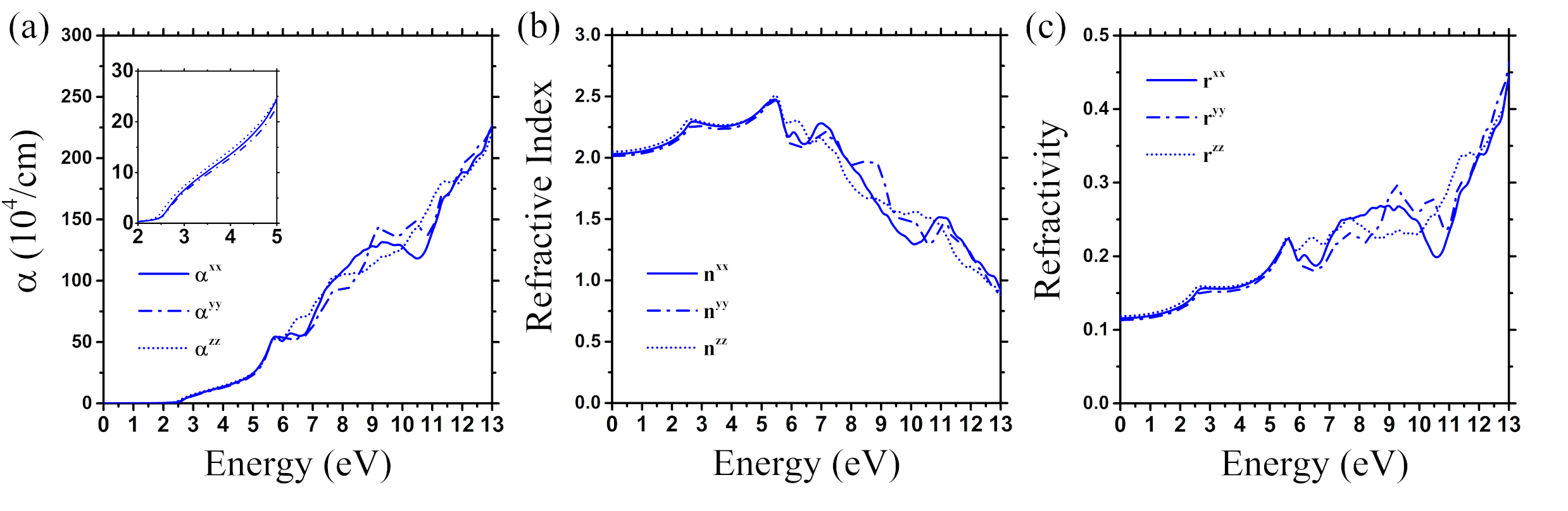}
\caption{
(a) The absorption coefficient $\alpha(\omega)$, (b) the refractive index n($\omega$) and (c) the optical reflectivity r($\omega$) of MgSnN$_{2}$ compound as a function of energy for the means diagonal components.
\label{optic}
}
\end{figure*}

Our calculations of the absorption, $\alpha(\omega)$, for the studied compound indicate that the material shows significant optical absorption over a wide energy range. Figure~\ref{optic}(a) displays the calculated $\alpha(\omega)$ for the three diagonal components, which are in excellent agreement with spectroscopic ellipsometry measurements~\cite{greenaway2020}. 
Similar to the behavior of $\epsilon(\omega)$, $\alpha(\omega)$ exhibits isotropy at low energies and anisotropy at higher energies.

The absorption starts at the fundamental bandgap energy, arising from transitions at the $\Gamma$ point, where N $p$-states in the upper valence band transition to unoccupied Sn $s$-states/N $p$-states in the conduction band.
As energy increases, absorption also increases, corresponding to a minimum in $\epsilon_{1}(\omega)$.
The MgSnN$_{2}$ compound exhibits strong absorption, approximately $10^{5}$~cm$^{-1}$, in the visible light range.
However, the highest absorption peak occurs in the UV range.

The refractive index spectrum, $n(\omega)$, is shown in Fig.~\ref{optic}(b).
At zero frequency, the refractive index is $2.05$, which closely matches the experimentally reported values of $1.9$–$2.3$~\cite{greenaway2020} and $1.9$–$2.1$~\cite{fahad22}.

The surface behavior of the material is characterized by the optical reflectivity, $r(\omega)$, which is the ratio of incident flux to reflected flux.
The calculated $r(\omega)$ is shown in Fig.~\ref{optic}(c) as a function of energy.
At zero frequency, the reflectivity is less than $0.12\%$.
MgSnN$_{2}$ exhibits low reflectivity, approximately $0.15\%$, in the visible light range.
These results suggest that MgSnN$_{2}$ could be an efficient material for multi-junction solar cells and optoelectronic devices, such as active layers in LEDs.

\begin{table*}[!t]
\caption{\label{tab4} 
The calculated SLME solar cell parameters of MgSnN$_{2}$ for $f_{g} = 1$, $x = 2$~$\mu$m and at different temperatures are based on DFT absorption calculation using the mBJ method.
}
\begin{ruledtabular}
\begin{tabular}{ccccccccc}
$T$ (K) & $J_{sc}$ ($A/m^{2}$) & $V_{oc}$ (V) & $J_{max}$ ($A/m^{2}$) & $V_{max}$ (V) & $P_{max}$ ($W/m^{2}$)& $FF$ & $\eta$ ($\%$)\\
\hline
273.15 & 67.2 & 2.13 & 66.7 & 2.02 & 134.7 & 0.941 & 13.45 \\
298.15 & 67.2 & 2.09 & 66.5 & 1.98 & 131.7 & 0.937 & 13.17 \\
323.15 & 67.2 & 2.06 & 66.2 & 1.95 & 129.2 & 0.933 & 12.90 \\
343.15 & 67.2 & 2.04 & 66.1 & 1.92 & 126.9 & 0.925 & 12.68 \\
\end{tabular}
\end{ruledtabular}
\end{table*}

\subsection{Photovoltaic properties}

\begin{figure}[!b]
\centering
\hspace*{-0.5 cm} 
\includegraphics[width=1.1\linewidth]{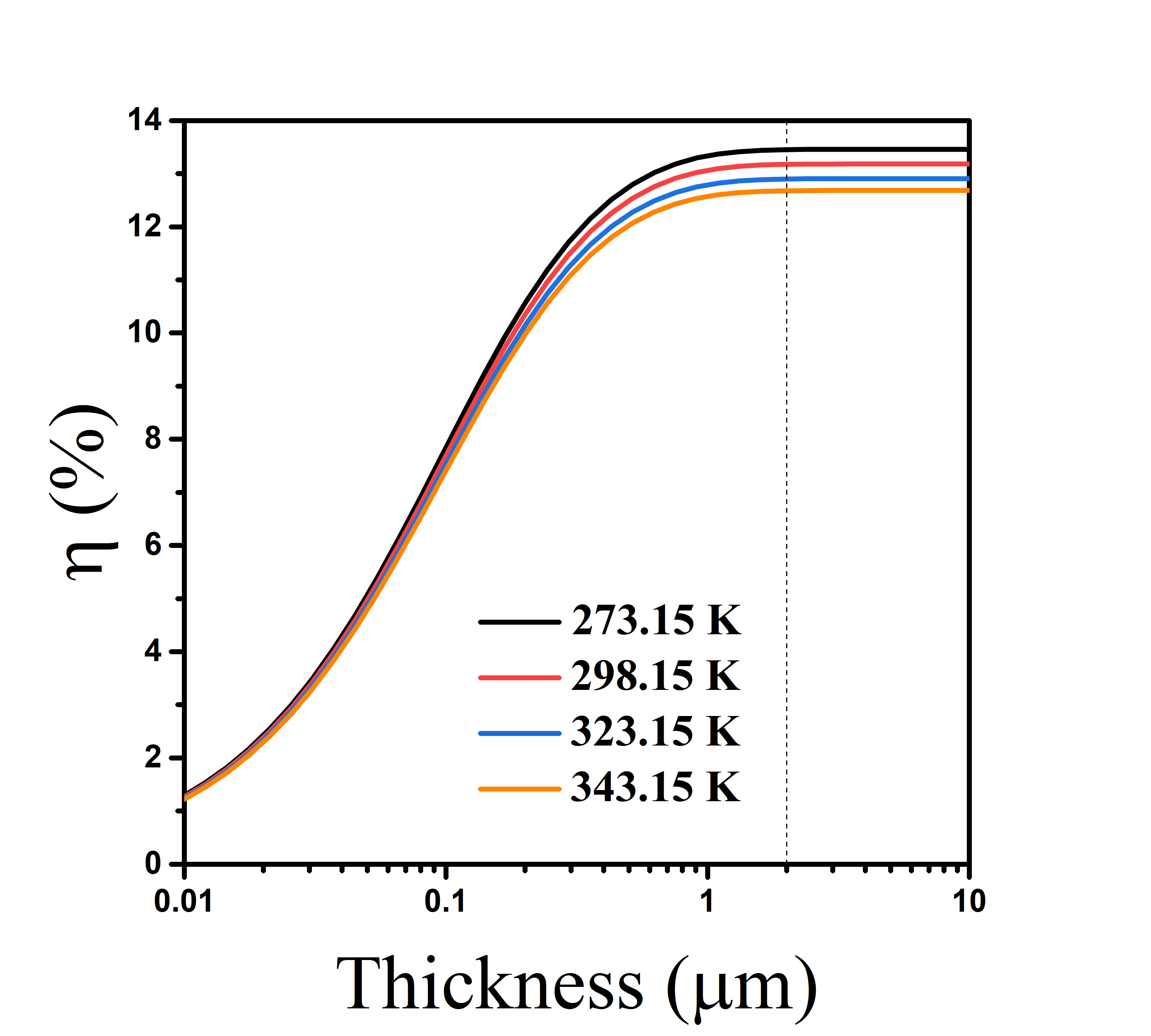}
\caption{SLME photovoltaic maximum efficiencies of MgSnN$_{2}$ as a function of film thickness at different temperatures. The dashed line represents a thickness of $2~\mu$m.
\label{effe}
}
\end{figure}

To conduct a comprehensive analysis of SLME, we first examine the theoretical efficiency, as defined in Eq.(\ref{TE}), which has been calculated as a function of absorber layer thickness and is presented in Fig.\ref{effe}.
The efficiency increases with absorber thickness due to enhanced photon absorption.
At greater thicknesses, the efficiency saturates because, for a sufficiently thick absorbing layer, the absorptivity approaches an ideal step function. At a thickness of $x = 2~\mu$m, the efficiency reaches a maximum of $13.17\%$ at room temperature. However, as the temperature increases, the efficiency declines due to a corresponding reduction in power density, primarily driven by the temperature-dependent decrease in the maximum voltage, $V_{max}$.

\begin{figure}[!t]
\centering
\hspace*{-0.5 cm}
\includegraphics[width=1.1\linewidth]{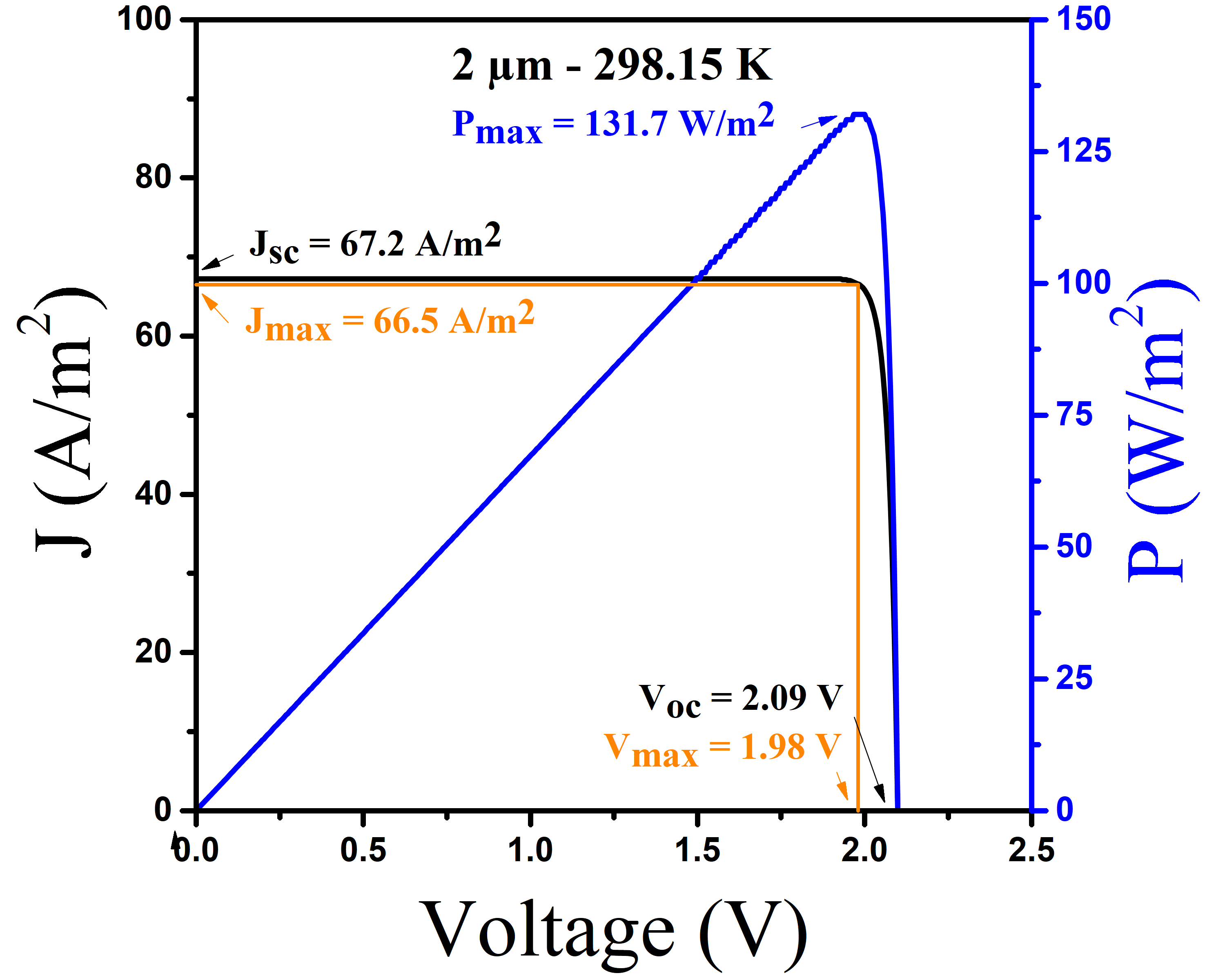}
\caption{The SLME current-voltage (black curve) and power--voltage (blue curve) characteristics  at room temperature for $f_{g} = 1$ and $x = 2$~$\mu$m of MgSnN$_{2}$. The orange lines indicate the maximum output current density, $J_{max}$, and voltage, $V_{max}$, corresponding to the maximum power density, $P_{max}$.
\label{SLME}
}
\end{figure}

Furthermore, the current density-voltage, $J$–$V$, and power density-voltage, $P$–$V$, characteristics have been calculated and plotted in Fig.\ref{SLME} for a thickness of $x = 2~\mu$m at room temperature.
Our calculations reveal that a 2~$\mu$m-thick MgSnN$_2$ film exhibits a high open-circuit voltage, $V_{oc}$, of 2.09~V.
This is attributed to its wide bandgap, consistent with the well-established correlation between high $V_{oc}$ and large bandgap semiconductors~\cite{yin15}.
The direct wide bandgap of MgSnN$_{2}$ minimizes radiative recombination current density, $J_0$, enhancing $V_{oc}$.
However, this also restricts the short-circuit current density, $J_{sc}$, due to its absorption edge lying in the UV range, which limits infrared and visible light absorption.
Consequently, $J_{sc}$ is calculated to be 67.2~A/m$^2$ at room temperature.
The close proximity of $J_{sc}$ and $J_{max}$ values indicates efficient charge collection, while the minimal voltage loss between $V_{oc}$ (2.09~V) and $V_{max}$ (1.98~V) suggests a high-quality junction.
The maximum output power density, $P_{max}$, and maximum efficiency are determined to be 131.7~W/m$^2$ and 13.17$\%$, respectively.
Furthermore, the high fill factor of 0.937 underscores the strong potential of MgSnN$_2$ for high-efficiency solar cell applications, driven by the minimal losses in current and voltage.

The influence of temperature, $T$, on the critical solar cell parameters is summarized in Table~\ref{tab4}.
From the data, it is evident that the power density decreases as the temperature increases, at an average rate of $2.66$~W/m$^{2}$ per $25$~K (or $^{\circ}$C) temperature increment.
This decline in power density is attributed to the reduction in maximum voltage, $V_{max}$, as temperature rises. 
The decrease in open-circuit voltage, $V_{oc}$, is primarily caused by an increase in electron-hole recombination rates within the solar cell at elevated temperatures, which diminishes its efficiency.
The short-circuit current density, $J_{sc}$, on the other hand, remains unaffected by temperature variations, as it primarily depends on the incident solar radiation rather than thermal effects.

\begin{figure}[!b]
\centering
\hspace*{-0.5 cm}
\includegraphics[width=\linewidth]{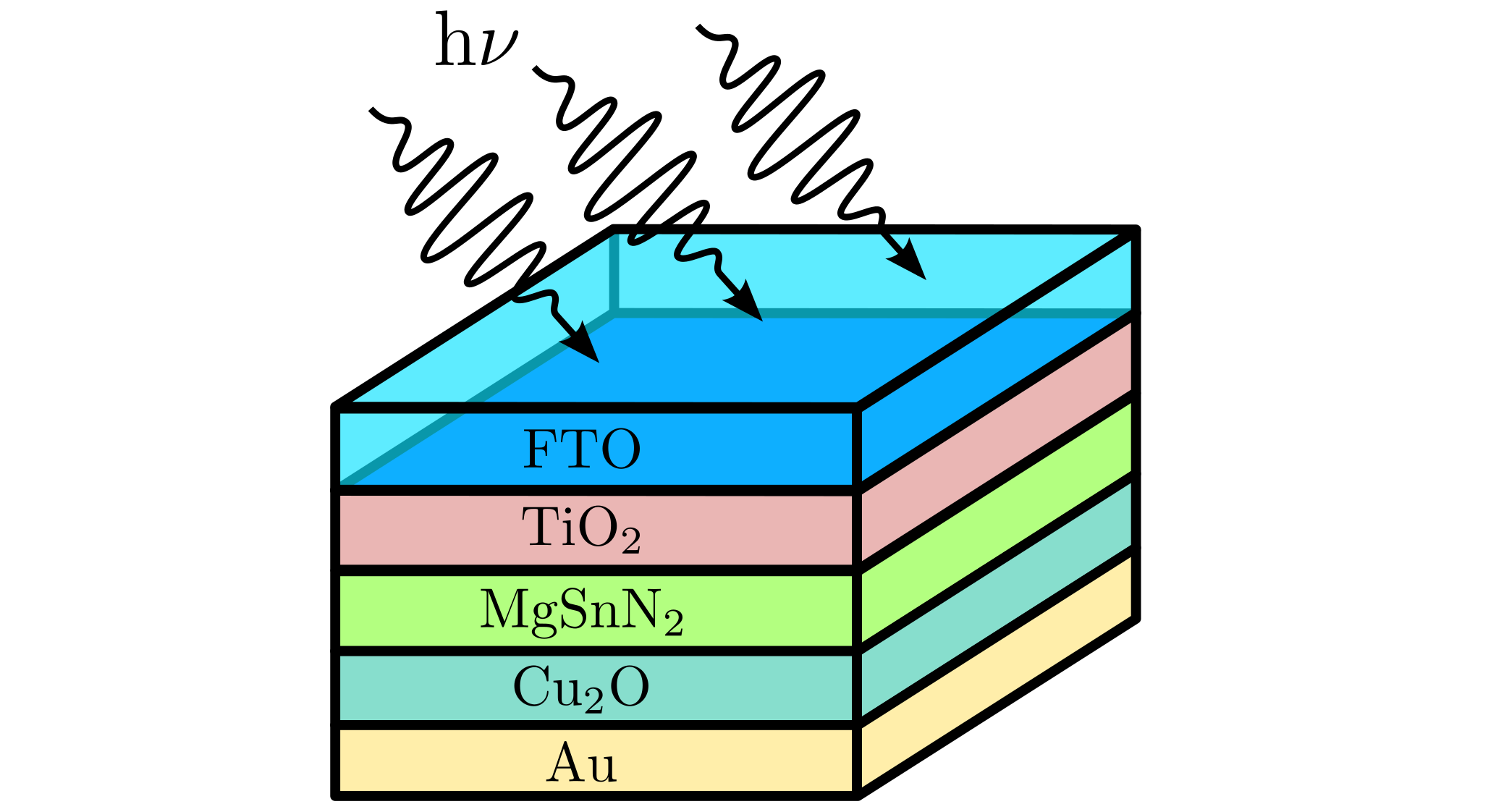}
\caption{Diagrammatic design configurations of an FTO/TiO$_{2}$/MgSnN$_{2}$/Cu$_{2}$O/Au solar cell.
\label{cell}
}
\end{figure}

Finally, to evaluate our proposed MgSnN$_{2}$ for multi-junction applications, we fabricated a single-junction device to validate our SLME results.
Here, a numerical simulation using the one-dimensional Solar Capacitance Simulator SCAPS-1D software package~\cite{scaps} is used to model and simulate a MgSnN$_{2}$ based solar cell.
The fundamental equations are Poisson’s equation and the continuity equations for electrons and holes~\cite{raoui2019}.
The heterovalent ternary nitride material MgSnN$_{2}$ is constructed on a Fluorine-doped Tin Oxide (FTO) substrate, with $n$-type TiO$_{2}$ used as the Electron Transport Layer (ETL), while the $p$-type Cu$_{2}$O is serving as the Hole Transport Layer (HTL). 
The gold (Au) is used as the back contact. 
The configuration of the FTO/TiO$_{2}$/MgSnN$_{2}$/Cu$_{2}$O/Au solar cell is illustrated in Fig.~\ref{cell}.

\begin{table*}[!t]
\caption{\label{tab5} 
Material parameters for layers of FTO/TiO$_{2}$/MgSnN$_{2}$/TiO$_{2}$/CuInS$_2$/Cu$_{2}$O/Au  tandem solar cell~\cite{issam24, rutlege13, hao2014, abega21, kale21, hossain22, shang14, bonnet13, sherkar17}.
}
\begin{ruledtabular}
\begin{tabular}{cccccc}
 & FTO & TiO$_{2}$ (ETL) & MgSnN$_{2}$ & CuInS$_2$ & Cu$_{2}$O (HTL) \\
\hline
Thickness ($\mu$m) & 0.5 & 0.023 & 2 & 0.08 & 0.24 \\
$Eg$ (eV) & 3.5 & 3.2 & 2.456 & 1.216 & 2.2 \\
Electron affinity (eV) & 4 & 4.26 & 4 & 4.5 & 3.2 \\
Relative permittivity & 9 & 9 & 4.05 & 5.73 & 7.1 \\
CB effective DOS $N_{c}$ (1/cm$^{3}$) & 2.2 $\times$ 10$^{18}$ & 2 $\times$ 10$^{18}$ & 2.92 $\times$ 10$^{18}$ & 2.2 $\times$ 10$^{18}$ & 10$^{18}$ \\
VB effective DOS $N_{v}$ (1/cm$^{3}$) & 1.8 $\times$ 10$^{19}$ & 1.8 $\times$ 10$^{19}$ & 2.74 $\times$ 10$^{18}$ & 1.8 $\times$ 10$^{19}$ & 10$^{18}$ \\
Electron thermal velocity (cm/s) & 10$^{7}$ & 10$^{7}$ & 2.37 $\times$ 10$^{7}$ & 10$^{7}$ & 10$^{7}$ \\
Hole thermal velocity (cm/s) & 10$^{7}$ & 10$^{7}$ & 2.42 $\times$ 10$^{7}$ & 10$^{7}$ & 10$^{7}$ \\
Electron mobility (cm$^{2}$/V.s) & 20 & 20 & 73.2 & 100 & 2 $\times$ 10$^{2}$ \\
Hole mobility (cm$^{2}$/V.s) & 20 & 10 & 76.4 & 25 & 86 $\times$ 10$^{2}$ \\
Donor density $N_{D}$ (1/cm$^{3}$) & 2 $\times$ 10$^{19}$ & 9 $\times$ 10$^{16}$ & 0 & 0 & 0 \\
Acceptor density $N_{A}$ (1/cm$^{3}$) & 0 & 0 & 10$^{15}$ & 4 $\times$ 10$^{15}$ & 10$^{18}$  \\
Density of defects & 10$^{15}$ & 10$^{15}$ & 10$^{14}$ & 10$^{14}$ & 10$^{15}$ \\
\end{tabular}
\end{ruledtabular}
\end{table*}

\begin{table*}[!t]
\caption{\label{tab6} 
Interfacial defect layers parameters~\cite{issam24, rutlege13, hao2014, abega21, kale21, hossain22, shang14, bonnet13, sherkar17}.}
\begin{ruledtabular}
\begin{tabular}{cccccc}
 & FTO/TiO$_{2}$ & TiO$_{2}$/MgSnN$_{2}$ & MgSnN$_{2}$/Cu$_{2}$O & TiO$_{2}$/CuInS$_{2}$ & CuInS$_{2}$/Cu$_{2}$O\\
\hline
Defect type  & Neutral & Neutral & Neutral & Neutral & Neutral\\
Capture cross section for electron (cm$^{3}$) & 10$^{-19}$ & 10$^{-19}$ & 10$^{-19}$ & 10$^{-19}$ & 10$^{-19}$\\
Capture cross section for hole   & 10$^{-19}$ & 10$^{-19}$ & 10$^{-19}$ & 10$^{-19}$ & 10$^{-19}$\\
Energetic distribution & Single & Single & Single & Single & Single\\
Energy level with respect to $E_v$ (eV) & 0.6 & 0.6 & 0.6 & 0.6 & 0.6\\
Total density & 10$^{15}$ & 10$^{15}$ & 10$^{15}$ & 10$^{15}$ & 10$^{15}$\\
\end{tabular}
\end{ruledtabular}
\end{table*}

The electrical parameters used in the simulations are outlined in Table~\ref{tab5}, with references to pertinent computational and experimental literature. 
The parameters considered are film thickness, which is a primary factor in solar cell performance, relative dielectric constant, $\epsilon_{r}$, bandgap, E$_g$, electron affinity, $\chi$, effective state densities of the conduction and valence bands, $N_c$ and $N_v$, respectively, electron and hole mobilities, $\mu_n$ and $\mu_p$, respectively, electron and hole thermal velocities, $v_n$ and $v_p$, respectively, donor and acceptor densities, $N_D$ and $N_A$, respectively, and total defect density, $N_t$. 
In addition, Table~\ref{tab6} details the interface qualities of FTO/TiO$_{2}$, TiO$_{2}$/MgSnN$_{2}$, and MgSnN$_{2}$/Cu$_{2}$O, allowing an assessment of the defect concentration.

The thickness of the absorber is a critical factor in determining the performance of solar cell devices, and it is necessary to select the optimal thickness for maximum performance. 
Based on our prior SLME analysis, a thickness of 2~$\mu$m has been determined to be the optimal thickness for our absorbent layer to achieve high power output. 
The current density $J_{sc}$ increases in proportion to the absorber thickness, indicative of an associated escalation in the rate at which electron-hole pairs are generated within the absorber. 
When the thickness of the absorber layer increases from 1~$\mu$m to 2~$\mu$m, $J_{sc}$ increases from 95.7~A/m$^2$ to 113.23~A/m$^2$. 
The ETL, HTL and FTO are fixed at 0.023~$\mu$m, 0.24~$\mu$m, and 0.5~$\mu$m, respectively, and these thicknesses provide high efficiency for this solar cell, while maintaining reasonable cost.

The relative dielectric permittivity of the material under investigation is 4.05, which corresponds to the dielectric permittivity (the imaginary part of the dielectric function) of an incident photon energy of 0 eV. 
For FTO, TiO$_{2}$, and Cu$_{2}$O, the electron affinities are 4~eV, 4.26~eV, and 3.2~eV, respectively. 
The electron affinity is a crucial metric for assessing the electron injection properties of materials, making it a pivotal factor in optoelectronic and solar cell applications.

The effective state densities of the conduction and valence bands, $N_c$ and $N_v$, are contingent on both temperature and the effective mass of both electrons and holes,  $m^{*}_{e,h}$. 
The formula for the effective density of states at the edges of the conduction and valence bands, $N_c$ and $N_v$, is given as follows~\cite{kasap2006}:
\begin{equation}
\label{Ncv}
N_{c,v} = 2g_{d} \left( \frac{2\pi m^{*}_{e,h} k_{B}T}{h^2} \right)^{3/2}.
\end{equation}

All simulations are conducted at 298.15~K in order to align with the results from the SLME analysis and AM1.5G solar spectra with 1000~W/m$^2$ power density. 
The degeneracy factor, $g_{d}$, is assumed to be 1 for purposes of simplicity in this simulation. The effective masses employed in this simulation are 0.24$m_{0}$ and 0.23$m_{0}$ for the electron and hole, respectively, where $m_{0}$ is the absolute mass of the electron~\cite{lambrecht2016, lambrecht2019}.

\begin{table*}[!t]
\caption{\label{tab7} 
SCAPS-1D photovoltaic parameters of FTO/TiO$_{2}$/MgSnN$_{2}$/Cu$_{2}$O/Au single-junction solar cell.
}
\begin{ruledtabular}
\begin{tabular}{ccccccccc}
$T$ (K) & $J_{sc}$ ($A/m^{2}$) & $V_{oc}$ (V) & $J_{max}$ ($A/m^{2}$) & $V_{max}$ (V) & $P_{max}$ ($W/m^{2}$)& $FF$ & $\eta$ ($\%$)\\
\hline
273.15 & 113.23 & 1.349 & 110.62 & 1.160 & 128.38 & 0.840 & 12.83 \\
298.15 & 113.23 & 1.348 & 110.43 & 1.159 & 128.36 & 0.838 & 12.80 \\
323.15 & 113.23 & 1.348 & 110.20 & 1.155 & 127.33 & 0.834 & 12.73 \\
343.15 & 113.23 & 1.347 & 110 & 1.151 & 126.62 & 0.830 & 12.66 \\
\end{tabular}
\end{ruledtabular}
\end{table*}

\begin{table*}[!t]
\caption{\label{tab8} 
SCAPS-1D photovoltaic parameters of FTO/TiO$_{2}$/MgSnN$_{2}$/TiO$_{2}$/CuInS$_{2}$/Cu$_{2}$O/Au tandem solar cell at room temperature.
}
\begin{ruledtabular}
\begin{tabular}{ccccccccc}
$T$ (K) & $J_{sc}$ ($A/m^{2}$) & $V_{oc}$ (V) & $J_{max}$ ($A/m^{2}$) & $V_{max}$ (V) & $P_{max}$ ($W/m^{2}$)& $FF$ & $\eta$ ($\%$)\\
\hline
Top & 113.23 & 1.348 & 110.43 & 1.159 & 128.36 & 0.838 & 12.80 \\
Bottom (under AM1.5G) & 174.09 & 0.899 & 167.45 & 0.801 & 134.127 & 0.857 & 13.41 \\
Bottom (under filtered spectra) & 113.7 & 0.888 & 109.52 & 0.791 & 86.63 & 0.858 & 8.66 \\
Tandem & 113.23 & 2.245 & 111.547 & 2.010 & 224.21 & 0.882 & 22.42 \\
\end{tabular}
\end{ruledtabular}
\end{table*}

Thermal velocities, $v^{n,p}_{th}$, play a significant role in a number of processes, including carrier transport, recombination and generation. 
Thermal velocities affect Shockley-Read-Hall (SRH) recombination, where the carrier lifetime, $\tau$, is inversely proportional to $v^{n,p}_{th}$, thus influencing charge trapping and recombination rates. 
Higher thermal velocities result in faster carrier interactions, increasing recombination and decreasing carrier lifetime, whereas lower values improve charge transfer. 
Thermal velocity also affects carrier mobility and diffusion, which in turn affects device performance. 
The thermal velocities of the absorber under consideration are determined using Eq.(\ref{vt})~\cite{kasap2006}, while the $v^{n,p}_{th}$ of the other layers are derived from other studies:
\begin{equation}
\label{vt}
v^{n,p}_{th} = \sqrt{\frac{3k_{B}T}{m^{*}_{e,h}}}.
\end{equation}

Electron and hole mobilities, $\mu_n$ and $\mu_p$, respectively, are pivotal parameters that affect charge transport, recombination and overall device performance. 
Mobility determines how fast charge carriers move under an electric field, it is related to the relaxation time, $\tau$, by $\mu_{(n,p)} = q\tau/m^{*}_{(e,h)}$.
Higher mobility improves carrier transport and reduces recombination losses, thereby enhancing solar cell efficiency. 
The electron and hole mobilities calculated for MgSnN$_2$ are presented in Table~\ref{tab5}.

In conclusion, the SCAPS-1D simulations of our single-junction solar cell yield an efficiency of up to $12.80\%$ at 298.15~K, which closely aligns with the SLME calculation of $13.17\%$.
The minor discrepancy arises from electronic and optical losses, as well as interfacial defects in various layers.
Table~\ref{tab7} summarizes the impact of temperature on key solar cell parameters.
The large bandgap of MgSnN$_2$ results in limited absorption within the UV range, thereby, the fabrication of multi-junction devices is advantageous in this instance.
However, it should be noted that multi-junction (tandem) solar cell devices are not fully supported by SCAPS-1D software.
Consequently, the top and bottom devices are simulated individually, and the tandem solar cell is fabricated using two sub-cells connected in series.
The MgSnN$_2$-based solar cell functions as the top sub-cell in the tandem device, while a ternary chalcopyrite CuInS$_2$ serves as the light harvester in the bottom sub-cell.
The bottom layer utilizes similar HTL and ETL, as well as interfacial defects parameters [see Table~\ref{tab6}.]

\begin{figure*}[!t]
\centering
\includegraphics[width=\linewidth]{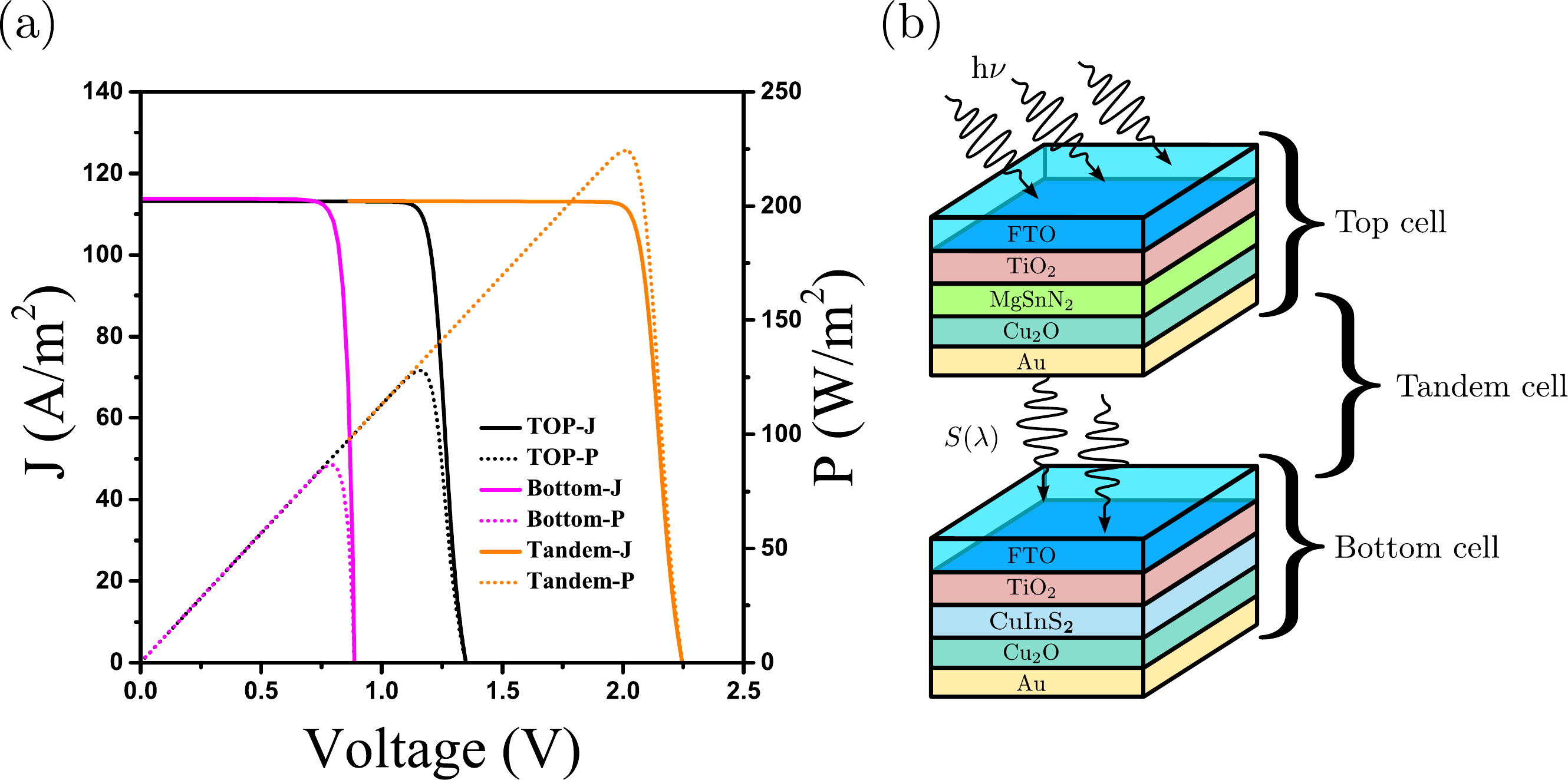}
\caption{(a) $J-V$ (line) and $P-V$ (dotted) curves of the top, bottom sub-cells, and tandem device at room temperature. (b) Diagrammatic design configurations of an FTO/TiO$_2$/MgSnN$_2$/Cu$_2$O/TiO$_2$/CuInS$_2$/Cu$_2$O/Au tandem solar cell.
\label{tandem}
}
\end{figure*}

The top sub-cell receives light directly from the sun (AM1.5G spectra), whereas the bottom sub-cell obtains the illumination from the filtered spectrum through the top sub-cell.
The filtered spectrum is generated by applying the following equation:
\begin{equation}
\label{vt}
S(\lambda) = S_0(\lambda)\exp\left(\sum_{i=1}^{4}-\alpha_i(\lambda)d_i\right).
\end{equation}
Where, $S_0(\lambda)$ denotes the incident light spectrum, $\alpha_i$ and $d_i$ represent the absorption coefficient and thickness of each layer ($i$) of the top sub-cell, respectively. 
The reflection losses from each interface are not considered. 
Fig.~\ref{tandem}(b) represents a diagram of the tandem configuration. 
The top and bottom sub-cells are connected in series; hence, similar currents should travel through the two sub-cells.
Furthermore, unmatched short-circuit currents result in the concentration of charge carriers at the recombination contact, which can increase recombination losses and reduce overall efficiency.
This is due to the fact that excess carriers from the higher-current cell have no path to flow, resulting in recombination at the interface rather than contributing to power generation~\cite{bonnet13, sherkar17}. 
To address this, the thickness of CuInS$_2$ is varied to obtain similar short-circuit currents between the top and bottom sub-cells.
The multi-junction device voltage is determined by the summation of the voltages of the two sub-cells.

The $J-V$ and $P-V$ curves for the top sub-cell, the bottom sub-cell and the tandem device are depicted in Fig. \ref{tandem}(a).
The bottom sub-cell, the filtered spectra and the top sub-cell demonstrate similar short-current densities, thus yielding a comparable short-circuit current density for the multi-junction device. 
The lower open-circuit voltage of the bottom sub-cell compared to the top sub-cell is primarily attributable to the difference in bandgaps of CuInS$_2$ (1.216 eV) and MgSnN$_2$ (2.45 eV) and the discrepancy in their electrical properties.
Despite the lower bandgap of CuInS$_2$, the power output and efficiency of the bottom sub-cell under filtered spectra are lower than those of the top sub-cell, as demonstrated in Table~\ref{tab8}, this decrease is mainly due to the smaller thickness of the absorber layer in the bottom sub-cell compared to the top sub-cell, the decrease in open circuit voltage, in addition to the attenuation of the light spectra after passing through the top sub-cell.

An improvement in the open-circuit voltage of the tandem device is evident following the integration of the two sub-cells, resulting in an elevated maximum voltage, power output density, efficiency and fill factor.
The efficiency of the single-junction solar cell was found to increase from 12.80$\%$ to 22.42$\%$ after the fabrication of the multi-junction device through the association of the top and the bottom sub-cells in series.

\section{Summary}
\label{sec.sum}

The orthorhombic crystal structure of MgSnN$_{2}$, belonging to the $Pna2_1$ space group, can be derived from the wurtzite structure by substituting two group III atoms with one group II atom and one group IV atom. 
The calculated lattice constants and wurtzite-like parameters, obtained using LDA and GGA-PBE methods, align well with previous theoretical findings.
MgSnN$_{2}$ is a direct bandgap semiconductor at the $\Gamma$ point, featuring an isotropic conduction band edge and an anisotropic valence band edge. 
This reflects isotropic electron effective masses and anisotropic hole effective masses. 
Bandgap calculations reveal that the mBJ method predicts a higher bandgap of $2.45$~eV, compared to $1.26$~eV and $1.31$~eV obtained using LDA and GGA, respectively.
The total density of states (TDOS) analysis highlights $p$-$p$ coupling between N $p$-states and Sn $p$-states at the valence band maximum (VBM), while the conduction band minimum (CBM) comprises a mix of Sn $s$-states and N $p$-states, resulting in $sp^3$ hybridization characteristic of tetrahedral coordination.
Optical property calculations demonstrate that MgSnN$_{2}$ achieves its highest absorption peak in the ultraviolet (UV) range, underscoring its strong potential for UV-based applications. 
Furthermore, it exhibits storng absorption in the visible spectrum.
Coupled with a low reflectivity, MgSnN$_{2}$ emerges as a promising candidate for use as an absorber layer in tandem solar cells targeting the high-energy region of the visible spectrum.
SLME photovoltaic performance analysis suggests that a $2$~$\mu$m-thick MgSnN$_{2}$ film can generate a power density of $131.7$~W/m$^2$ and maximum efficiency of $13.17$\% at room temperature accompanied by an excellent fill factor, $FF$, of $0.937$.
This efficiency has been further validated through the fabrication of a single-junction device.

Moreover, a multi-junction device was simulated using two sub-cells connected in series.
The simulation shows expected improvements in output parameters such as open-circuit voltage, maximum voltage, and power density. The efficiency of the tandem device increases from 12.80$\%$ (single junction) to 22.42$\%$.
These characteristics position MgSnN$_{2}$ as a promising candidate for the top sub-cell in multi-junction solar architectures.
Introduction of cation disorder has been shown to reduce the bandgap~\cite{makin2019} and enhance low-energy absorption~\cite{greenaway2020}, enabling fine-tuning of optoelectronic properties and further improving its potential for solar cell applications.

\begin{acknowledgments}
Some figures in this work were rendered using {\sc Vesta}~\cite{momma.izumi.11}.
We kindly acknowledge support by National Science Centre (NCN, Poland) under Project No.~2021/43/B/ST3/02166.
\end{acknowledgments}

\bibliography{biblio.bib}


\end{document}